%% file: aspect-check.tex
\documentclass[10pt,a4paper]{book}
\usepackage{perugia}
\begin{document}

\input{marcucci.tex}

\end{document}

%% file: marcucci.tex
\talktitle{An application of the Independent Component Analysis
methodology to gamma ray astrophysical imaging}

\talkauthors{Francesca Marcucci \structure{a,b},
             Claudia Cecchi \structure{a,b},
             Gino Tosti  \structure{a,b}}

\authorstucture[a]{Dipartimento di Fisica,
                   Universit\`a di Perugia,
                   via Pascoli, 06100~Perugia, Italy}

\authorstucture[b]{INFN, Sezione di Perugia,
                   via Pascoli, 06100~Perugia, Italy}

\shorttitle{ICA method: application to simulated images}

\firstauthor{F. Marcucci, C. Cecchi, G. Tosti}

\begin{abstract}
Independent Component Analysis (ICA) \cite{ICA} is a statistical
method often used to decompose a complex dataset in its
independent sub-parts. It is a powerful technique to solve a
typical Blind Source Separation problem. A fast calculation of the
gamma ray sky observed by GLAST, assuming the expected
instrumental response, has been implemented. The simulated images
were used to test the capability of the ICA method in identifying
the sources.
\end{abstract}
\section{Introduction}
Maps produced in large area surveys contain a linear mixture of
signals from several astrophysical and cosmological sources
convoluted with the spatial and spectral response of the detector.
GLAST, thanks to its sensitivity (about 50 times better than EGRET
at $E>100$ MeV) and its wide field of view (more than $2.5 sr$),
will reveal a large number of sources during its lifetime, giving
scientists a superior data sample to study many interesting
phenomena.

A common problem encountered is to detect and analyze these
sources. Many classical tools (i.e. likelihood) need assumptions
on a theoretical model in order to fit the data and to extract
physical parameters.

To optimize the computation time and to minimize the data
modelling assumptions, an alternative technique based on
Independent Component Analysis has been investigated. It could be
a valid tool to identify signals of different origin in the sky
maps. A previous application of ICA in the astrophysical domain is
due to Baccigalupi \emph{et al.} \cite{Baccigalupi}, to separate
the contribution of Cosmic Microwave Background in Plank simulated
images.

In section 2 we will outline the ICA methodology to extract an
independent component from a set of linearly mixed sources, while
in section 3 a preliminary study of the ICA performance on gamma
ray source spatial information will be described.

\section{ICA for astrophysical images}
The method requires that the independent components are
statistical independence and at most one of them is gaussian. For
astrophysical images the previous conditions are theoretically
satisfied and also the linear model holds exactly.

Let us denote with $\bf{s \equiv (s_{1},...,s_{m})^{T}}$ the
vector of the M signal sources, where each $\bf{s_{i}}$ is the
individual image of the i-th source (with the total T pixels
stacked row by row into a T-vector), and with $\bf{x \equiv
(x_{1},...,x_{n})^{T}}$ the vector of the observed signal in N
different energy bands, where each $\bf{x_{i}}$ is a T-vector as
above. The analytical form for the ICA model is $x=As$,
 where $\bf{A}$ is a mixing matrix to be determined. Using
the central limit theorem, ICA is able to solve the problem. Let
us define a matrix $\bf{W}$ such that the transformed $\bf{y}$
vectors are as independent as possible
\begin{equation}\label{eqn:ica}
\hat{s}=y=Wx
\end{equation}
We can derive the independent component $\bf{s}$ by minimizing the
dependence of the $\bf{y}$ vectors, that means, for the central
limit theorem, to maximize the non gaussianity of the $\bf{Wx}$
vector. The FastICA \cite{FastICA} algorithm allows to find a
matrix $\bf{W}$ as the best estimator of the inverse matrix
$\bf{A^{-1}}$, minimizing the negentropy function, that gives a
quantitative measure of the gaussianity, for the $\bf{y}$
variable. The columns of the $\bf{W}$ matrix are updated by the
iteration:
\begin{equation}\label{eqn:fastica}
  w'=E\{xg(w^{T}x)\}-E\{g'(w^{T}x)\}w
\end{equation}
and then the matrix $\bf{W}$ is orthonormalized. The function $g$
in equation (\ref{eqn:fastica}) is an odd nonlinear function and
$g'$ its derivative, we assume $g(u)=\tanh(u)$. Once the algorithm
has converged, the estimation of the single components can be
obtained using equation (\ref{eqn:ica}).

\section{FastICA and simulated GLAST maps}
In order to test the capability of the ICA method in identifying
the dominant contributions in an astrophysical map, images of a
chosen region of the sky, as observed by the GLAST telescope, have
been simulated using the GLAST Light Simulator program described
in \cite{light sim}. Maps of $41\times41$ or $21\times21$
(T-dimension=1681 or 441) pixels ($1 px=0.5°$) in seven different
energy ranges (N-dimension=7), between 10 MeV and 1 GeV, have been
generated. The independent simulated components included in the
maps are: the diffuse background (galactic and extragalactic), the
faint sources generated randomly and the sources from the Third
Egret Catalogue.

The performance of the algorithm for different signal (sources) to
noise (diffuse background) conditions have been tested.

\begin{figure}[!h]
\begin{minipage}[!h]{0.5\linewidth}
\includegraphics[scale=0.3]{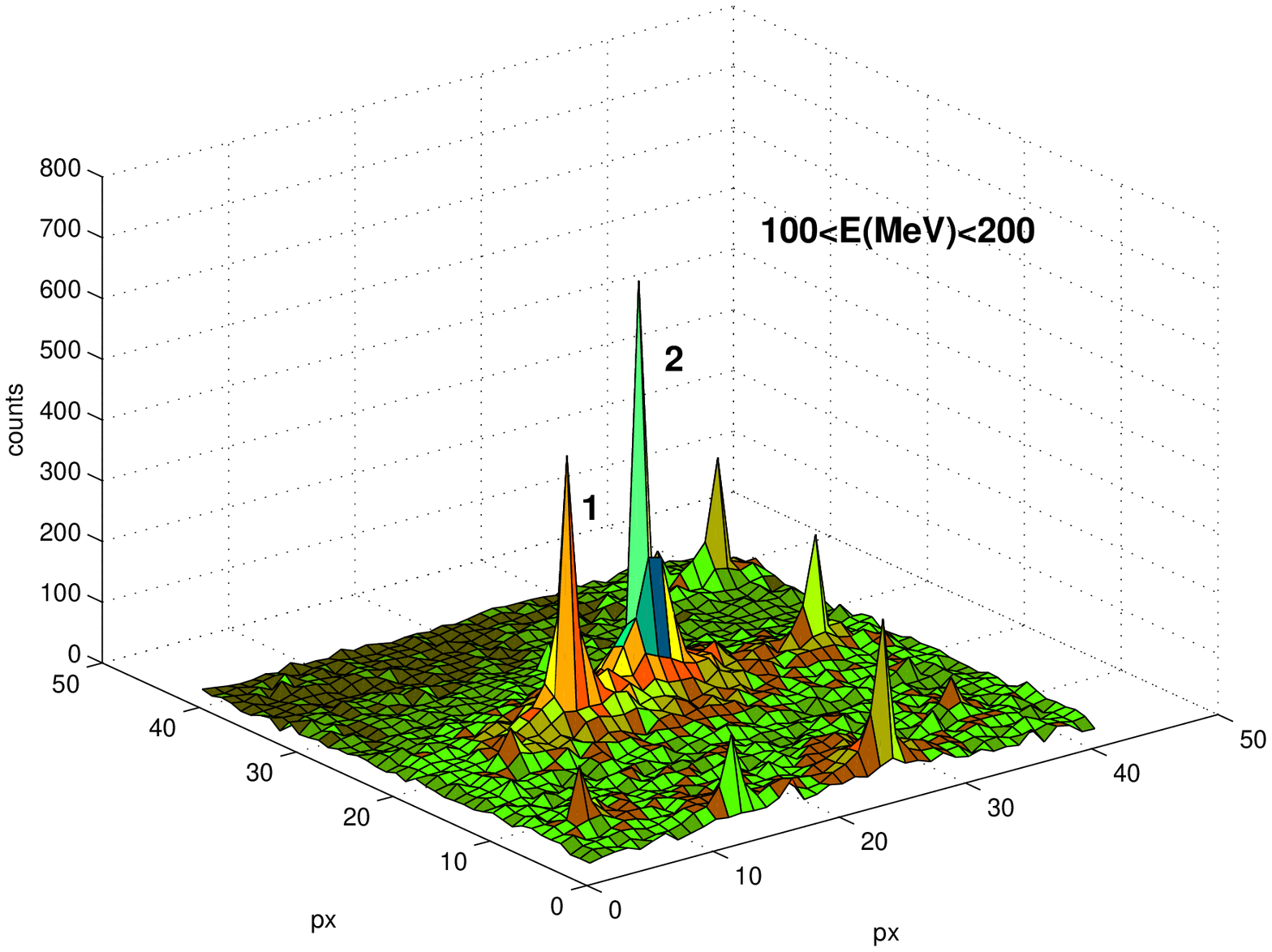}
\caption{One of the input images around 3C279 and $\sim 54$ day of
exposure.}\label{fig:first input}
\end{minipage}
\hskip 0.3 cm
 \begin{minipage}[!h]{0.45\linewidth}
\includegraphics[scale=0.3]{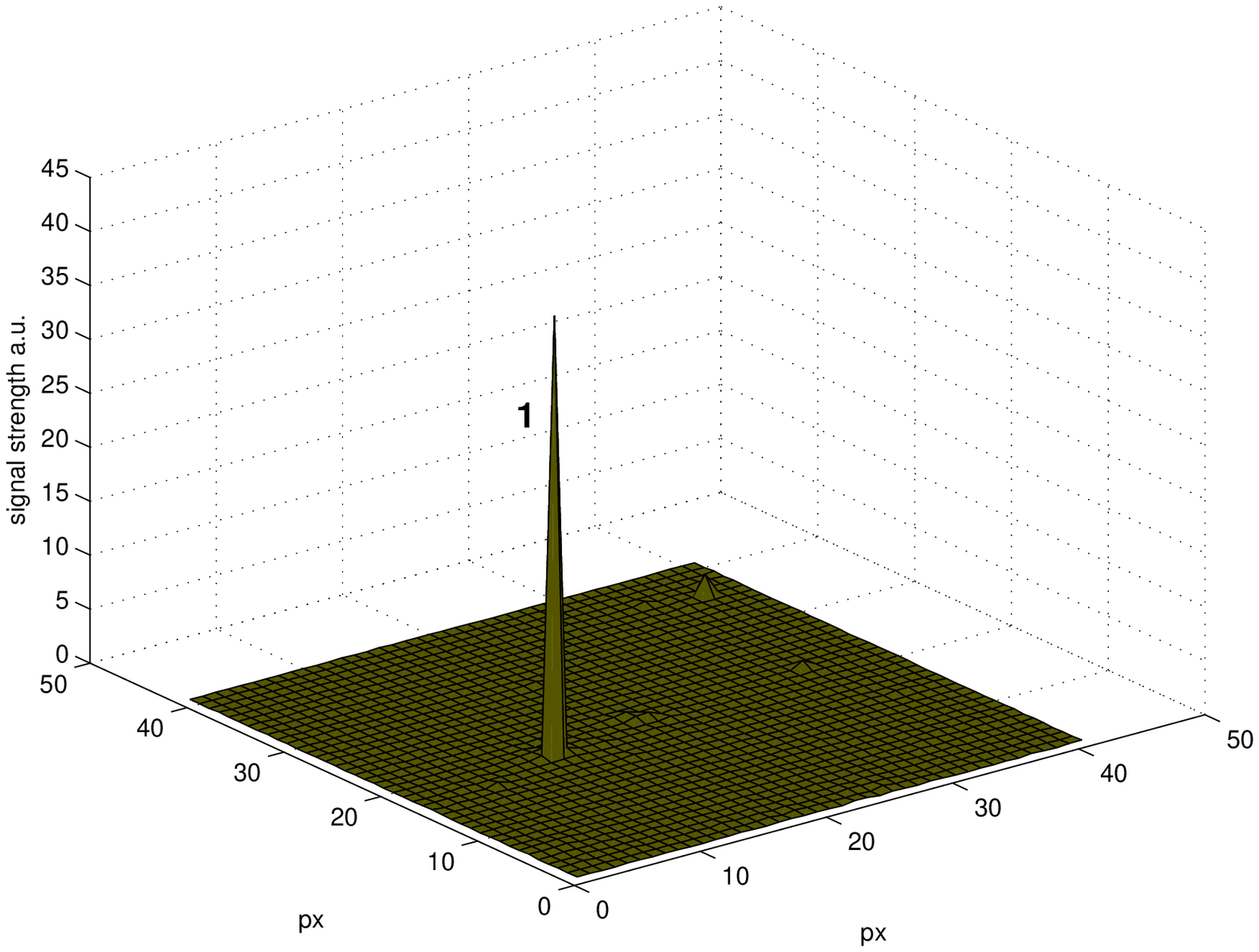}
\caption{One of the output images: the source 1 has been
identified}\label{fig:first output}
\end{minipage}
\end{figure}

\begin{figure}[!h]
\begin{minipage}[!h]{0.5\linewidth}
\includegraphics[scale=0.3]{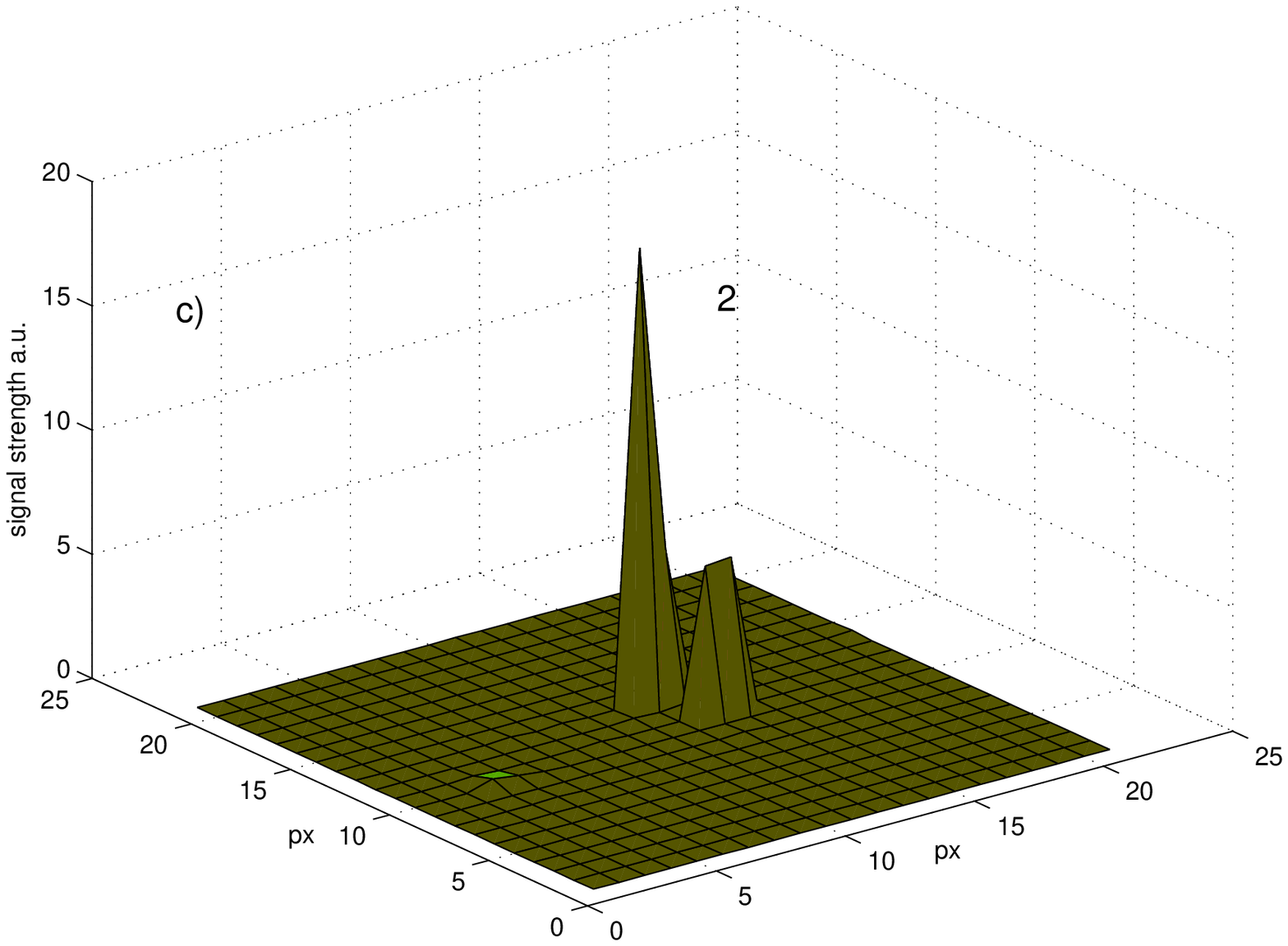}
\caption{Number of photons from the sources reduced by a factor
150. One of the output images: source 2 has been
identified}\label{fig:second output}
\end{minipage}
\hskip 0.3 cm
 \begin{minipage}[!h]{0.45\linewidth}
\includegraphics[scale=0.3]{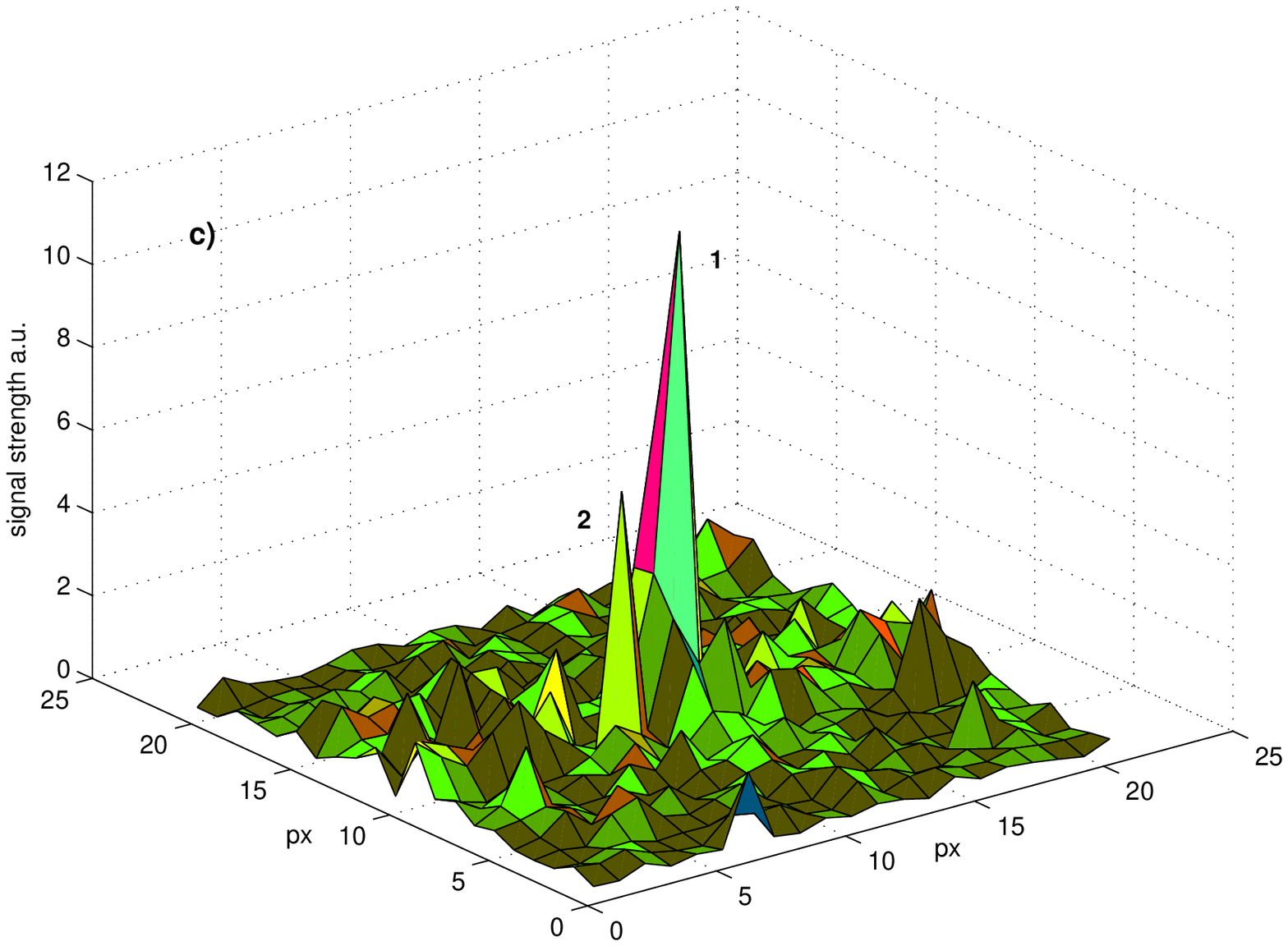}
\caption{Region around the Galactic Center. One of the output
images: sources 1 and 2 are identified}\label{fig:third output}
\end{minipage}
\end{figure}

First images in a sky region around blazar 3C279 with a long
exposure period (one precession period $\sim 54$ days) have been
simulated; in this region there is a large contribution from the
sources (fig.\ref{fig:first input}) because we are far from the
galactic plane, where the background is dominant. The method works
correctly and the position of the sources in the region are
correctly reconstructed (fig.\ref{fig:first output}).
Reducing the windows to $21\times21$ pixels, results on
information also about the faint sources, even if their
contribution is very small because of the presence of the brighter
ones. Subtracting the contribution of latter would allow this
technique to identify sources in the shadow. Looking at the same
region, after reducing the number of photons from the sources by a
factor 150, in order to have comparable contributions from signal
and noise, the method is always able to extract the dominant
components (fig.\ref{fig:second output}).

In a second step, the observation windows has been moved towards
the galactic center and an exposure of about 10 days has been
simulated. In this region the contribution from the diffuse
background is considerable larger, nevertheless the method is able
to correctly reconstruct the position of the sources
(fig.\ref{fig:third output}).

In the previous cases the method identifies the sources but
sometimes some of them are reconstructed together as a single
component, especially if they have peaks with very low intensity.
The following test case has been studied to test the capability of
ICA to separate sources in presence of a superposition. In the
input images the distance between the peaks of the sources has
been reduced to about 1 pixel. As shown in fig.\ref{fig:last
output}, each source is distinguished in a component, despite of
the small distance between them in the input maps.

In this first application of the ICA method, information about the
signal intensity have not been carried out, but only indications
about the position of the sources have been studied.
\begin{figure}[!h]
\begin{center}
\includegraphics[scale=0.3]{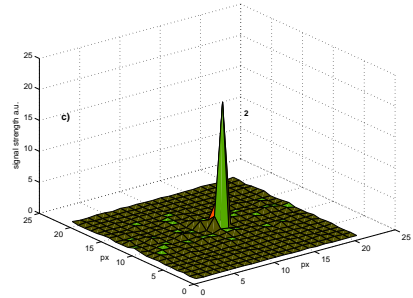}
\caption{Superposition of the sources. One of the output images:
source 2 has been identified.}\label{fig:last output}
\end{center}
\end{figure}
\vspace{-0.6cm}
\section{Conclusions}
A preliminary application of the FastICA algorithm to test the
capability of the method in a source localization problem has been
described. Under different signal to noise conditions it works
properly giving results according to the simulated inputs. The
method has been applied only to the reconstruction of the sources
position; its effectiveness on the absolute intensity will be
further investigated.